\magnification=\magstep1
\overfullrule=0pt
\parskip=6pt
\baselineskip=15pt
\headline={\ifnum\pageno>1 \hss \number\pageno\ \hss \else\hfill \fi}
\pageno=1
\nopagenumbers
\hbadness=1000000
\vbadness=1000000

\vskip 25mm
\vskip 25mm
\vskip 25mm

\centerline{\bf SPECIAL WEIGHTS AND ROOTS FOR FINITE LIE ALGEBRAS}
\vskip 15mm

\centerline{\bf H. R. Karadayi}
\centerline{Dept. Physics, Fac. Science, Istanbul Tech. Univ.}
\centerline{80626, Maslak, Istanbul, Turkey }
\centerline{e-mail: karadayi@itu.edu.tr}

\vskip 5mm

\centerline{\bf M. Gungormez}
\centerline{Dept. Physics, Fac. Science, Istanbul Tech. Univ.}
\centerline{80626, Maslak, Istanbul, Turkey }
\centerline{e-mail: gungorm@itu.edu.tr}

\vskip 5mm
\medskip

\centerline{\bf{Abstract}}

With the introduction of {\bf special roots}, we show the existence of some
{\bf special weights} with quite interesting properties for finite Lie algebras.
We propose and discuss two statements which lead us to an explicit construction
of these special weights and roots. It is seen that the special weights provide 
us a basis to express Weyl group actions directly no matter what the Weyl group
elements would be.


\vskip 15mm
\vskip 15mm
\medskip

\hfill\eject

\vskip 3mm
\noindent {\bf{I.\ DEFINITION OF SPECIAL WEIGHTS AND ROOTS }}
\vskip 3mm

In the Cartan-Weyl tradition of a finite Lie algebra ${\bf {\cal L}_r}$ of rank  r ,
{\bf simple roots} $\alpha_i$'s and their duals, {\bf the fundamental dominant weights},
$\lambda_i$'s play a principal role. Their very definitions and also some of their
properties will be given here briefly with reference of the excellent book of
Humphreys{\bf [1]}. Throughout the work, we always assume $i,j,k = 1,2, \dots, r$
and $A,B = 1,2, \dots dimW({\bf {\cal L}_r})$ where $W({\bf {\cal L}_r})$ is the 
{\bf Weyl Group} of Lie algebra $ {\bf {\cal L}_r} $.

Let  ${\bf {\cal C}_r}$ be Cartan-Matrix of  ${\bf {\cal L}_r}$. The simple roots can
then be thought of as r-dimensional linear vectors satisfying scalar products
$$ <\alpha_i,\alpha_j> = ({\bf {\cal C}_r})_{i,j} .  \eqno(I.1)$$
Within the scope of this work, the {\bf Root Lattice} ${\bf {\cal R}_r}$ and the
{\bf positive root lattice} ${\bf {\cal R}_r}^+$ will be considered respectively
as is composed out of integral and nonnegative-integral linear superpositions of these
simple roots. The {\bf Weight Lattice}  ${\bf {\cal W}_r}$ is thus defined to be the 
dual of root lattice. For any two elements $ \lambda , \alpha \in {\bf {\cal W}_r} $,
the existence of a non-symmetric scalar product
$$ < \lambda , \alpha > \equiv 2 \ { (\lambda,\alpha) \over (\alpha,\alpha) } $$
is always guaranteed by (I.1) where $ (\lambda,\alpha) $ is the symmetrical scalar
product which is known to exist for any two r-dimensional vectors. The set
$$ {\cal H}(1) \equiv \{ \lambda_1, \lambda_2, \dots, \lambda_r \}  $$
of fundamental dominant weights are then defined by
$$ <\lambda_i,\alpha_j> \equiv \delta_{i,j} \ \ . \eqno(I.2) $$
Now we look for some other sets
$$ {\cal H}(A) \equiv \{ \Lambda_A(1), \Lambda_A(2), \dots, \Lambda_A(r) \} \ \ , \ \
A=1,2, \dots, D  \eqno(I.3) $$
with the property that
$$ (\lambda_i,\lambda_j) = (\Lambda_A(i),\Lambda_A(j)) . \eqno(I.4) $$
Elements $\Lambda_A(i)$ of these sets will then be called {\bf special weights}.
Note here that (I.4) is central in the definition of special weights. The maximal value D of the number of sets ${\cal H}(A)$ has also some implications as will be
seen in the following.

To find the special weights explicitly, we need the following two conjectures which
lead us in fact the definition of special roots in a natural way.

\noindent $ {\underline {\bf Conjecture \ 1}}$

\noindent Let $\Gamma(i)^+$ be the subset of ${\bf {\cal R}_r}^+ $ defined by
$$\Gamma(i)^+ \equiv \{ \alpha \in {\bf {\cal R}_r}^+   \ / \
<\lambda_i,\alpha> = 1 \} \ \ . \eqno(I.5)  $$
We then have
$$ dim\Gamma(i)^+ = dimW(\lambda_i) $$
for $i=1,2, \dots ,r$.

It is important to notice here that no element of $\Gamma(i)^+ $ could be equal to an
element of the Weyl orbit $W(\lambda_i)$ of $\lambda_i$. Special weights can then be
specified explicitly as a result of our second conjecture:

\noindent $ {\underline {\bf Conjecture \ 2}}$

\noindent Any element $\Lambda_A(i) \in {\cal H}(A)$ has the form
$$ \Lambda_A(i) = \lambda_i - \gamma_A(i) \eqno(I.6)  $$
where $\gamma_A(i) \in \Gamma(i)^+$ and
$$ D = dimW({\bf {\cal L}_r}) \ \ .  \eqno(I.7)   $$

Note here, due to notation, that the possibilities  $\gamma_A(i) = \gamma_B(i)$ or
$\gamma_A(i) = \gamma_A(j)$ could be valid for some A,B or i,j. Then, it is natural
to call $\gamma_A(i)$'s encountered in the explicit forms of special weights to be
{\bf special roots} which are in fact elements of the sets
$$ {\cal R}(A) \equiv \{ \gamma_A(1), \gamma_A(2), \dots, \gamma_A(r) \} \ \ , \ \
A=1,2, \dots, D \ \ . \eqno(I.8) $$

We are now ready to state why special weights are so special. Our statement for
$\Sigma_A \in W({\bf {\cal L}_r})$ here is simply that
$$ \Sigma_A(\lambda_i) = \Lambda_A(i) \ \ . \eqno(I.9) $$

We must emphasize here that $\Sigma_A$'s are Weyl group elements whereas 
$\Sigma_A(\lambda_i)$'s give us their action on the weight lattice. For an explicit
construction of Weyl group elements, one of the best that we can do is to find a 
presentation of Weyl group in terms of which its elements are to be expressed as 
multiple products of simple reflections, the elements corresponding to simple roots. 
It is well known however that,except some quite simple ones, this is already what one 
can not do in almost all cases for which the right hand side of (I.9) proves useful.

In the next section, we will explicitly show how our two conjectures hold in general for

$A_r$ Lie algebras. Beside a previous article {\bf[2]}, a detailed discussion will be given
for $G_2$ in another work {\bf[3]} which is complementary to this one. The case for $F_4$ 
will also be presented by one of the authors separately {\bf[4]}. Explicit tabulations of special roots are also given for $E_6$ and $E_7$ {\bf[5]}. As a nice application of (I.9), 
we will show, in a subsequent paper {\bf[6]}, that Weyl character formula can be applied explicitly in almost all cases with only one exception of $E_8$ Lie algebra for which we actually need some other tools in explicit calculations {\bf[7]}.

\vskip 3mm
\noindent {\bf{II.\ $A_r$ LIE ALGEBRAS }}
\vskip 3mm

In this section, we show how one lets our two conjectures to be true for $A_r$ Lie
algebras. By presenting $A_3$ Weyl group elements in terms of multiple products
of its simple reflections we also show that (I.9) is in fact true.

To show the validity of conjecture 1, we seek, for each and every $ \lambda_k$, the
solutions
$$ \gamma = m_1 \alpha_1 + m_2 \alpha_2 + \dots  + m_r \alpha_r  \eqno(II.1) $$
of the equation
$$  (\lambda_k,\gamma) = {1 \over 2} \ (\gamma,\gamma) \ \ . \eqno(II.2) $$
which is in fact equivalent to
$$ m_1^2 + m_2^2 + \dots m_r^2 = m_1 m_2 + m_2 m_3 + \dots +m_{r-1} m_r + m_k 
\eqno(II.3) $$
where $m_k$'s are non-negative integers. This simply shows that the solutions (II.1) 
are restricted by conditions
$$ m_1, m_2, \dots, m_r \leq k  \ \ , \ \ k=1,2, \dots, {r \over 2} \ ( {r+1 \over 2} )  $$
depending on whether r is an even (odd) integer. Let us note here that the solutions are
related for $\Gamma(k)^+$ and $\Gamma(r+1-k)^+$ due to diagram automorphisms of $A_r$ Lie
algebras. In result, we can say that we only have a finite number of solutions for any
$\Gamma(k)^+$.

To expose these solutions, let us introduce another fundamental system of weights $\mu_K$
defined by
$$ \alpha_k \equiv \mu_k - \mu_{k+1}   \eqno(II.4) $$
together with the condition
$$ \mu_1 + \mu_2 + \dots + \mu_{r+1} \equiv 0 \ \  .  \eqno(II.5)  $$
Being in coincidence with our earlier works, we would like to call $\mu_K$'s
{\bf fundamental weights} {\bf [8]} for K=1,2, \dots, r+1 . The complete set 
of solutions (II.1) can then be expressed by
$$ \Gamma(k)^+ = \bigcup_{s=0}^k \ \Gamma^{(s)}(k) \ \ . \eqno(II.6)  $$
Beside the set $ \Gamma^{(0)}(k) = \{ 0 \} $ which consists of trivial solution 
$\gamma = 0$, any partial set of solutions $\Gamma^{(s)}(k)$ here is supposed 
to have the form
$$ \Gamma^{(s)}(k) \equiv \{ \ \mu_{i_1} + \mu_{i_2} + \dots + \mu_{i_s} - (
\mu_{j_1} + \mu_{j_2} + \dots + \mu_{j_s} ) \}  \eqno(II.7)  $$
providing
$$ i_1,i_2, \dots, i_s = 1,2, \dots, k \ \ , \ \ j_1,j_2, \dots, j_s=k+1,k+2, \dots, r+1 \ \ . \eqno(II.8)  $$
It is important to notice here that no any two indices can take the same value in (II.7).
One thus has 
$$ dim \Gamma^{(s)}(k) = B[k,s] \ B[r+1-k,s]  $$
for which 
$$ B[n,m] \equiv {n! \over (n-m)! \ m! } $$
is the Binomial coefficient.
We therefore totally have
$$ 1 + \sum_{s=1}^k \ B[k,s] \ B[r+1-k,s] = B[r+1,k] $$
number of solutions. Our conjecture is now clear for $A_r$ Lie algebras in view of the
fact that $ dimW(\lambda_k) = B[r+1,k] $.

To show that the conjecture 2 is also holds here, we remark that (I.4) is equivalent to
$$ (\gamma_A(i),\gamma_A(j)) = {s_i + s_j \over 2}  $$
where  $\gamma_A(i) \in \Gamma^{(s_i)}(i) $, $\gamma_A(j) \in \Gamma^{(s_j)}(j) $
and $ s_i, s_j = 1,2, \dots ,k $ for $\Gamma(k)^+$, as in (II.6).

\hfill\eject

Let us illustrate this in the simple but non-trivial example of $A_3$. We first need an explicit presentation of $A_3$ Weyl group elements $\Sigma_A$'s for $A=1,2, \dots, 24.$ This will be given on the left-hand side of the following table in terms of multiple products of simple reflections $\sigma_i$ which correspond to simple roots $ \alpha_i$ . Its right-hand side gives us, on the othet hand, the sets of special roots by the aid of which we see that (I.9) is in fact valid for $A_3$ Lie algebra. 

$$ \vbox{\tabskip=1pt \offinterlineskip
\halign to380pt{\strut#& \vrule#\tabskip=0em plus2em& \hfil#&
\vrule#& \hfil#\hfil&
\vrule#\tabskip=0pt\cr \noalign{\hrule}
&& $\eqalign{
\Sigma_1&=1  \cr
\Sigma_2&=\sigma_1 \cr
\Sigma_3&=\sigma_2 \cr
\Sigma_4&=\sigma_3 \cr
\Sigma_5&=\sigma_1 \ \sigma_2   \cr
\Sigma_6&=\sigma_1 \ \sigma_3   \cr
\Sigma_7&=\sigma_2 \ \sigma_1   \cr
\Sigma_8&=\sigma_2 \ \sigma_3   \cr
\Sigma_9&=\sigma_3 \ \sigma_2   \cr
\Sigma_{10}&=\sigma_1 \ \sigma_2 \ \sigma_1  \cr
\Sigma_{11}&=\sigma_1 \ \sigma_2 \ \sigma_3  \cr
\Sigma_{12}&=\sigma_1 \ \sigma_3 \ \sigma_2  \cr
\Sigma_{13}&=\sigma_2 \ \sigma_1 \ \sigma_3  \cr
\Sigma_{14}&=\sigma_2 \ \sigma_3 \ \sigma_2  \cr
\Sigma_{15}&=\sigma_3 \ \sigma_2 \ \sigma_1  \cr
\Sigma_{16}&=\sigma_1 \ \sigma_2 \ \sigma_1 \ \sigma_3  \cr
\Sigma_{17}&=\sigma_1 \ \sigma_2 \ \sigma_3 \ \sigma_2  \cr
\Sigma_{18}&=\sigma_1 \ \sigma_3 \ \sigma_2 \ \sigma_1  \cr
\Sigma_{19}&=\sigma_2 \ \sigma_1 \ \sigma_3 \ \sigma_2  \cr
\Sigma_{20}&=\sigma_2 \ \sigma_3 \ \sigma_2 \ \sigma_1  \cr
\Sigma_{21}&=\sigma_1 \ \sigma_2 \ \sigma_1 \ \sigma_3 \ \sigma_2  \cr
\Sigma_{22}&=\sigma_1 \ \sigma_2 \ \sigma_3 \ \sigma_2 \ \sigma_1  \cr
\Sigma_{23}&=\sigma_2 \ \sigma_1 \ \sigma_3 \ \sigma_2 \ \sigma_1  \cr
\Sigma_{24}&=\sigma_1 \ \sigma_2 \ \sigma_1 \ \sigma_3 \ \sigma_2 \ \sigma_1  \cr} $
&& $\eqalign{
{\cal R}(1)&= \{0, 0, 0 \}   \cr
{\cal R}(2)&= \{\alpha_1, 0, 0 \}   \cr
{\cal R}(3)&= \{0, \alpha_2, 0 \}   \cr
{\cal R}(4)&= \{0, 0, \alpha_3 \}   \cr
{\cal R}(5)&= \{\alpha_1, \alpha_1+\alpha_2, 0 \}   \cr
{\cal R}(6)&= \{\alpha_1, 0, \alpha_3 \}   \cr
{\cal R}(7)&= \{\alpha_1+\alpha_2, \alpha_2, 0 \}   \cr
{\cal R}(8)&= \{0, \alpha_2, \alpha_2+\alpha_3 \}   \cr
{\cal R}(9)&= \{0, \alpha_2+\alpha_3, \alpha_3 \}   \cr
{\cal R}(10)&= \{\alpha_1+\alpha_2, \alpha_1+\alpha_2, 0 \}   \cr
{\cal R}(11)&= \{\alpha_1, \alpha_1+\alpha_2, \alpha_1+\alpha_2+\alpha_3 \}   \cr
{\cal R}(12)&= \{\alpha_1, \alpha_1+\alpha_2+\alpha_3, \alpha_3 \}   \cr
{\cal R}(13)&= \{\alpha_1+\alpha_2, \alpha_2, \alpha_2+\alpha_3 \}   \cr
{\cal R}(14)&= \{0, \alpha_2+\alpha_3, \alpha_2+\alpha_3 \}   \cr
{\cal R}(15)&= \{\alpha_1+\alpha_2+\alpha_3, \alpha_2+\alpha_3, \alpha_3 \}   \cr
{\cal R}(16)&= \{\alpha_1+\alpha_2, \alpha_1+\alpha_2, \alpha_1+\alpha_2+\alpha_3 \}   \cr
{\cal R}(17)&= \{\alpha_1, \alpha_1+\alpha_2+\alpha_3, \alpha_1+\alpha_2+\alpha_3 \}   \cr
{\cal R}(18)&= \{\alpha_1+\alpha_2+\alpha_3, \alpha_1+\alpha_2+\alpha_3, \alpha_3 \}   \cr
{\cal R}(19)&= \{\alpha_1+\alpha_2, \alpha_1+2 \ \alpha_2+\alpha_3, \alpha_2+\alpha_3 \}   \cr
{\cal R}(20)&= \{\alpha_1+\alpha_2+\alpha_3, \alpha_2+\alpha_3, \alpha_2+\alpha_3 \}   \cr
{\cal R}(21)&= \{\alpha_1+\alpha_2, \alpha_1+2 \ \alpha_2+\alpha_3, \alpha_1+\alpha_2+\alpha_3 \}   \cr
{\cal R}(22)&= \{\alpha_1+\alpha_2+\alpha_3, \alpha_1+\alpha_2+\alpha_3, \alpha_1+\alpha_2+\alpha_3 \}   \cr
{\cal R}(23)&= \{\alpha_1+\alpha_2+\alpha_3, \alpha_1+2 \ \alpha_2+\alpha_3, \alpha_2+\alpha_3 \}   \cr
{\cal R}(24)&= \{\alpha_1+\alpha_2+\alpha_3, \alpha_1+2 \ \alpha_2+\alpha_3, \alpha_1+\alpha_2+\alpha_3 \} \cr } $ & \cr  \noalign{\hrule}
}} $$

\hfill\eject

In the table above, it is seen that elements of sets ${\cal R}(A)$ are chosen, 
respectively among elements of the sets
$$ \eqalign{
\Gamma(1)^+ &= \{ 0,\alpha_1,\alpha_1+\alpha_2,\alpha_1+\alpha_2+\alpha_3 \}  \cr
\Gamma(2)^+ &= \{ 0,\alpha_2,\alpha_2+\alpha_3,\alpha_1+\alpha_2,\alpha_1+\alpha_2+\alpha_3,\alpha_1+2 \ \alpha_2+\alpha_3 \}  \cr
\Gamma(3)^+ &= \{ 0,\alpha_3,\alpha_2+\alpha_3,\alpha_1+\alpha_2+\alpha_3 \}   }  $$
which are defined as in the first conjecture.

\vskip3mm
\noindent{\bf {REFERENCES}}
\vskip3mm

\item [1] J.E.Humphreys, Introduction to Lie Algebras and Representation Theory, N.Y.,
\item \ \ \ \ Springer-Verlag (1972)
\item [2] H.R.Karadayi and M.Gungormez, On the Calculation of Group Characters, 
\item \ \ \ \ math-ph/0008014
\item [3] H.R.Karadayi and M.Gungormez, On an Explicit Calculation of Signatures (submitted)
\item [4] M.Gungormez, work in preparation
\item [5] http://atlas.cc.itu.edu.tr/ $\sim$ karadayi
\item [6] H.R.Karadayi and M.Gungormez, Further Exploration of Weyl Character Formula(submitted) 
\item [7] H.R.Karadayi and M.Gungormez, Summing over the Weyl Groups of $E_7$ and $E_8$,
\item \ \ \ \ math-ph/9812014
\item [8] H.R.Karadayi, Anatomy of Grand Unifying Groups I and II, ICTP preprints,
\item \ \ \ IC/81/213 and 224
\item \ \ \  H.R.Karadayi and M.Gungormez, J.Phys.A: Math.Gen., 32 (1999) 1701-1707

\end